\documentclass[12pt]{iopart}  

\usepackage{graphicx}
\usepackage{bm}
\usepackage{color}
\usepackage{soul}
\usepackage[square,sort&compress,numbers]{natbib}
\usepackage[bookmarks,colorlinks,allcolors=blue]{hyperref}
 
\begin{document} 

\title[Whistler instability for bi-Kappa electrons]{Towards a general
  quasi-linear approach for the instabilities of bi-Kappa
  plasmas. Whistler instability}

\author{ P.~S. Moya$^{1,2}$, M. Lazar$^{1,3}$, and S. Poedts$^{1,4}$} 
\address{$^{1}$Centre for mathematical Plasma Astrophysics, KU
  Leuven, Celestijnenlaan 200B, B-3001 Leuven, Belgium}
\address{$^{2}$Departamento de F\'{\i}sica, Facultad de Ciencias, Universidad de Chile, Santiago, Chile}
\address{$^{3}$Institut f\"{u}r Theoretische Physik, Lehrstuhl IV:
  Weltraum- und Astrophysik, Ruhr-Universit\"{a}t Bochum, D-44780
  Bochum, Germany}
\address{$^{4}$Institute of Physics, University of Maria Curie-Skłodowska, Lublin, Poland} 
  
\ead{pablo.moya@uchile.cl}

\begin{abstract} 
Kappa distributions are ubiquitous in space and astrophysical poorly
collisional plasmas, such as the solar wind, suggesting that
microscopic and macroscopic properties of these non-equilibrium
plasmas are highly conditioned by the wave-particle interactions. The
present work addresses the evolution of anisotropic bi-Kappa (or
bi-$\kappa$-)distributions of electrons triggering instabilities of
electromagnetic electron-cyclotron (or whistler) modes. The new
quasi-linear approach proposed here includes time variations of the
$\kappa$ parameter during the relaxation of temperature
anisotropy. Numerical results show that $\kappa$ may increase for
short interval of times at the beginning, but then decreases towards a
value lower than the initial one, while the plasma beta and
temperature anisotropy indicate a systematic relaxation. Our results
suggest that the electromagnetic turbulence plays an important role on
the suprathermalization of the plasma, ultimately lowering the
parameter $\kappa$. Even though the variation of $\kappa$ is in
general negative ($\Delta \kappa <0$) this variation seems to depend
of the initial conditions of anisotropic electrons, which can vary
very much in the inner heliosphere.
\end{abstract}
%
\vspace{2pc}
%
\submitto{\PPCF}
%
\maketitle
%

%
\section{Introduction} 
\label{sec:intro}

There is currently an increasing interest for understanding the
complex activity of the Sun, conditioning the space weather and with
direct or indirect consequences on planetary environments and our
terrestrial climate and technologies. The huge amount of kinetic
energy released by the solar outflows, from slow to fast winds, or
during coronal mass ejections (CMEs), accumulates in the solar wind
mainly as kinetic anisotropies of plasma particles, namely,
temperature anisotropies or self-focused beaming populations. At large
heliospheric distances, e.g., 1 Astronomical Unit (AU), near the
Earth’s orbit, the solar wind is a collisionless non-equilibrium
plasma. It is well-known that in astrophysical and space environments
when collisions are inefficient, plasma particle velocity distribution
function (VDF) easily develop non-thermal characteristics that can
provide the necessary free energy to excite self-generated
instabilities, so called micro-instabilities. The excitation and
relaxation of these instabilities play an important role, especially
at kinetic scales in space plasma systems, such as the solar
wind~\cite{kasper2002,marsch2006,bale2009,bruno2013} and the Earth's
magnetosphere~\cite{heinemann1999,matsumoto2006,espinoza2018}.

For a realistic characterization of the enhanced wave fluctuations
contributing to the relaxation process, the proposed kinetic
approaches need also a realistic representation of the particle VDFs,
according to the observations. Standard bi-Maxwellian model (i.e.,
Maxwellian distribution function with two temperatures) offers the
simplest way to model magnetized plasma populations with anisotropic
temperatures, $T_\parallel \ne T_\perp$, where $\parallel$ and $\perp$
define directions relative to the local magnetic field lines
(gyrotropic distributions). Different combinations between temperature
anisotropy and plasma beta ($\beta_\parallel = 8\pi n k_B T_\parallel
/ B^2_0$) (where $n$, $k_B$ and $B_0$ represent the plasma density,
the Boltzmann constant and the strength of the background magnetic
field, respectively) will lead to different conditions for various
kinetic instabilities. Anisotropic electrons with
$T_{\perp}/T_{\parallel} > 1$ trigger instabilities (predominantly
right-handed polarized for small angles of propagation)
\cite{gary1993,gary2006,Lazar2018,Lopez2020}, while the left-handed
parallel or oblique firehose instabilities are excited for
$T_{\perp}/T_{\parallel} +2/\beta_\parallel <
1$~\cite{camporeale2008,Lazar2009,vinas2015}.

Analogous to a bi-Maxwellian, the bi-Kappa model was introduced to
describe anisotropic distributions with suprathermal tails
\cite{summers1991}. Kappa models consistently reproduce the
distributions of plasma particles in the magnetosphere and solar wind,
i.e., electrons~\cite[see
  e.g.][]{olbert1968,Vasyliunas1968,pilipp1987,nieves-chinchilla2008,espinoza2018},
and for
ions~\cite[e.g.][]{lockwood1987,chotoo1998,chotoo2000,espinoza2018}.
High-energy (power-law) tails of Kappa distributions are enhanced by
the suprathermal populations, and can markedly modify the dispersion
and (in)stability conditions, at both electron and ion scales
\cite{lazar2011,dossantos2014,navarro2015,vinas2015,vinas2017,Lazar2018,Lopez2019}. The
analysis should take into account not only the parameters quantifying
the kinetic anisotropy but also the abundance of suprathermals, in
this case the $\kappa$ parameter.

Wave-particle interactions can alter the shape of the VDF and
therefore the macroscopic properties of the plasma through a complex
non-linear process. One traditional way to include non-linear effects
of kinetic instabilities is the quasi-linear (QL) or weak turbulence
theory~\cite{vedenov,bernstein1,kennel1966a}. QL approaches have been
developed for temperature anisotropy driven instabilities in both,
bi-Maxwellian and bi-Kappa distributed plasmas. In the solar wind
conditions QL relaxation of bi-Maxwellian anisotropic plasmas (with
$T_{\perp}/T_{\parallel}\neq 1$) show a good agreement, qualitatively
and quantitatively, with fully non-linear particle
simulations~\cite[see
  e.g.][]{moya2012,seough2013,moya2014,Yoon2017,Shaaban2019}. Extended
to bi-Kappa distributed plasmas the same QL procedure shows that
suprathermals can modify the time-scales of temperature anisotropy
instabilities, accelerating the excitation and relaxation processes
\cite[see e.g.][]{Lazar2017,Lazar2018,Lazar2019}. However, in this
case comparisons between QL results and the higher order non-linear
simulations, such as Vlasov-Maxwell solvers~\cite{Lazar2017} or
Particle-In-Cell (PIC) simulations \cite{Lazar2018} may not show the
same good agreement as in the case of their bi-Maxwellian
counterpart. \cite{Lazar2017,Lazar2018} suggest that such disagreement
is mainly owed to the fact that QL approaches for Kappa distributions
are constrained to keep $\kappa$ parameter fixed in time, whereas in
PIC or Vlasov-Maxwell simulations the same parameter appears to
evolve, an evolution that, by construction, cannot be achieved in a
constrained QL model.

In this paper we address this issue by proposing a new QL approach for
modeling the growth and saturation of instabilities and the relaxation
of bi-Kappa distributions. Besides the time variation of the main
moments quantifying the anisotropic temperature components,
$T_{\parallel, \perp}$, our approach also accounts for a temporal
variation of the $\kappa$ parameter. To do so, we consider the
electromagnetic electron-cyclotron (EMEC) instability, also known as
whistler instability, which is driven by anisotropic electrons with
$T_\perp > T_\parallel$. The basic dispersion relations obtained from
linear (Vlasov-Maxwell) theory are presented in section 2, while in
Section~\ref{sec:qlt} we derive the QL equations for the EMEC
instability, by including $\kappa \to \kappa (t)$ as a function of
time. In the present analysis we propose a zero-order approximation,
which assumes the quasi-thermal core dominating the dynamics and
evolving independently of suprathermals. In our model we consider that
all the electrons follow a bi-Kappa velocity distribution. However,
due to their high number density ($>$ 90\% of total density), and by
their kinetic energy density \cite{Lazar-etal-2020}, the core
electrons may dominate the dynamics. Therefore, here we build a zero
order approximation, assuming an evolution of the core decoupled from
that of suprathermals in the tails, and implicitly from the variation
of kappa index. Numerical results are presented and discussed in
detail in Section~\ref{sec:results}, analysing systematically
different cases and taking several selections for the initial values
of temperature anisotropy, plasma beta and $\kappa$. Finally, in
Section~\ref{sec:conclusions} we summarize our findings and discuss
the insights obtained, along with perspectives of better understanding
of the dynamics of collisionless plasma in the solar wind.

\section{Vlasov linear dispersion relation: Kappa vs. Maxwellian}
\label{sec:theory}
%
For our model we consider a quasi-neutral magnetized plasma composed
by electrons and ions, in which electrons follow a bi-Kappa velocity
distribution function (VDF) given by

\begin{equation}
f(v_\perp,v_\parallel)
=\frac{n_0}{\pi^{3/2}\theta_\perp^2\theta_\parallel}
\frac{\Gamma(\kappa)}{\kappa^{1/2}\Gamma(\kappa-1/2)}
\Biggl(1+\frac{v_\perp^2}{\kappa\theta_\perp^2}
+\frac{v_\parallel^2}{\kappa\theta_\parallel^2}\Biggr)^{-\kappa-1},
\label{eq:fkappa}    
\end{equation}
where $n_0$ is the total number density, and $\kappa$ is the power-law
exponent quantifying the abundance of suprathermal particles in the
high-energy tails (suprathermalization) of our (bi-)Kappa distribution
model. Velocity parameters $\theta_{\parallel,\perp}$ used for the
normalization of perpendicular and parallel components of the velocity
in Eq.~(\ref{eq:fkappa}) relate to the corresponding components
$T_{\parallel,\perp}$ of the (kinetic) temperature $T= (T_\parallel +
2\,T_\perp)/3$, defined as the second order moment of the bi-Kappa
distribution function.

\begin{equation}
T_\parallel=\frac{m_e\theta_\parallel^2}{2k_B } \frac{\kappa}{\kappa-3/2},\qquad  
T_\perp=\frac{m_e\theta_\perp^2}{2k_B } \frac{\kappa}{\kappa-3/2}\,, 
\label{eq:temps}
\end{equation}
where $k_B$ is the Boltzmann constant and $m_e$ the electron mass.

If $\kappa$ increases, lowering the high-energy tails, in the limit case of $\kappa \to \infty$ we approach the bi-Maxwellian (quasi-)thermal core of our bi-Kappa model~\cite{Lazar2015,vinas2015} 
\begin{equation}
f_M(v_\perp,v_\parallel)=\frac{1}{\pi^{3/2}\theta_\perp^2\theta_\parallel} \exp \Biggl(-\frac{v_\perp^2}{\theta_\perp^2} 
- \frac{v_\parallel^2}{\theta_\parallel^2}\Biggr) \label{eq:fM}
\end{equation}
with components of the core temperature, $\Theta = (\Theta_\parallel + 2\Theta_\perp)/3$, defined by
\begin{equation}
\Theta_\parallel=\frac{m_e\theta_\parallel^2}{2k_B } ,\qquad  
\Theta_\perp=\frac{m_e\theta_\perp^2}{2 k_B }. \label{eq:thetas}
\end{equation}
Thus, with Eq.~(\ref{eq:temps}) in (\ref{eq:thetas}) we find the relationship between the Kappa temperatures ($T$, $T_{\perp,\parallel}$) and the core temperatures ($\Theta$, $\Theta_{\perp,\parallel}$)
\begin{equation}
    T_\perp=\Theta_\perp\, \frac{\kappa}{\kappa-3/2},\qquad T_\parallel=\Theta_\parallel\, 
\frac{\kappa}{\kappa-3/2}, \qquad    T =  \Theta\, \frac{\kappa}{\kappa-3/2}, \label{eq:tthkappa}
\end{equation}
Eqs.~(\ref{eq:tthkappa}) suggest already that time variation of the
Kappa temperatures may actually involve time variations of the core
temperatures and the $\kappa$ parameter as well, as we will see
below. It is also important to mention that in the limit
$\kappa\to\infty$ we have $T\to\Theta$ (the most probable speed for
Kappa becomes thermal speed for
Maxwellian)~\cite{Vasyliunas1968,Lazar2015,Lazar2016}.

In the case of bi-Kappa VDFs, the Vlasov linear dispersion relation
for right-hand (RH) polarized waves propagating in the direction of
the background magnetic field $\mathbf{B_0}$ is given by~\cite[see
  e.g][and references therein]{vinas2015,Lazar2017}
\begin{equation}
\frac{c^2k_\parallel^2}{\omega_{pe}^2}=\left(\frac{T_\perp}{T_\parallel}-1\right)+\left[\frac{T_\perp}{T_\parallel}\,\omega
-\left(\frac{T_\perp}{T_\parallel}-1\right)|\Omega_e|\right] \frac{1}{k_\parallel\theta_\parallel}\,Z_\kappa
\left(\frac{\omega-|\Omega_e|}{k_\parallel\theta_\parallel}\right)\,, \label{eq:disprel}
\end{equation}
where $\omega$ and $k_\parallel$ are the wave frequency and
wavenumber, respectively, and $\omega_{pe} = \sqrt{4\pi n_0 e^2/m_e}$
and $\Omega_{e} = e B_0/m_e c$ are the electron plasma frequency and
gyrofrequency. Also, $e$ is the elementary charge, $c$ is the speed of
light, and
\begin{equation}
Z_\kappa(\xi)=\frac{1}{\pi^{1/2}\kappa^{1/2}} \frac{\Gamma(\kappa)}{\Gamma(\kappa-1/2)}\int_{-\infty}^\infty
dx\,\frac{(1+x^2/\kappa)^{-\kappa}}{x-\xi},\qquad {\rm Im}(\xi)>0 \label{eq:zkappa}
\end{equation}
is the Modified Dispersion
function~\cite{summers1991,HellbergMace2002,vinas2017}. It is
important to mention that in the Maxwellian limit $\kappa \to \infty$
$Z_k$ becomes the well-known Plasma Dispersion Function $Z$ defined by
Fried and Conte~\cite{zeta}.
\begin{equation}
    Z(\xi)=\frac{1}{\pi^{1/2}}\int_{-\infty}^\infty dx\,\frac{\exp(-x^2)}{x-\xi},\qquad {\rm Im}(\xi)>0. \label{eq:z}
\end{equation}
In the same limit $\kappa \to \infty$ Eq.~(\ref{eq:disprel}) becomes
specific for the bi-Maxwellian core, and will be use in the next to
describe the core contribution independent of suprathermals in the
high energy tails.

\begin{table}
    \caption{Electron plasma parameters used in the present study$^*$.\label{t1}}
    \begin{tabular}{@{}lcccc}
    \br
    Case & $A$ & $\beta_{c, \parallel}$ & $\kappa$ & $\beta_{\parallel}$  \\   
    \mr
                 & 4.0   & 0.05   & 3.0   & 0.10 \\   
        Case I   & 4.0   & 0.50   & 3.0   & 1.00 \\
                 & 4.0   & 5.00   & 3.0   & 10.0 \\ 
    \mr
                 & 4.0   & 0.50   & 1.6   & 8.00 \\   
        Case II  & 4.0   & 0.50   & 3.0   & 1.00 \\
                 & 4.0   & 0.50   & 8.0   & 0.62 \\ %
    \mr
                 & 2.0   & 0.50   & 3.0   & 1.00 \\   
        Case III & 4.0   & 0.50   & 3.0   & 1.00 \\
                 & 8.0   & 0.50   & 3.0   & 1.00 \\ %
    \br 
    \end{tabular}\\
    {\footnotesize $^*$In all cases we consider $\omega_{pe}/\Omega_e=20$, and cold ions.}
\end{table}
\normalsize
%

For temperature anisotropy 
\begin{equation}
A = \frac{T_\perp}{T_\parallel}  = \frac{\Theta_\perp}{\Theta_\parallel} > 1, \label{eq:aniso}
\end{equation}
the dispersion relation Eq.~(\ref{eq:disprel}) is unstable to whistler
electron-cyclotron electromagnetic (EMEC) waves, with a maximum growth
rate that increases with increasing temperature anisotropy $A>1$ and
increasing parallel plasma beta $\beta_\parallel = 8\pi n_0 k_B
T_\parallel/B^2_0$~\cite[see e.g.][and references
  therein]{gary1993}. As mentioned in Section~\ref{sec:intro}, this
behaviour is similar for any collisionless plasma composed by
anisotropic electrons, and in the case of $\kappa$-distributed
electrons, several studies have been presented in the last few
years~\cite{navarro2015,vinas2015,Lazar2017,Lazar2018}. Here we
consider three cases with different combinations of temperature
anisotropy, plasma beta and the $\kappa$ parameter as shown in
Table~\ref{t1}. To make a proper comparison between Kappa and its
Maxwellian core (subscript $c$), as a limit case $\kappa \to \infty$,
we need to differentiate between their beta parameters
\begin{equation}
\beta_{\parallel} = \frac{8 \pi n_0 k_B T_\parallel}{B^2_0} = \frac{8 \pi n_0 k_B \Theta_\parallel}{B^2_0}\,\left(\frac{\kappa}{\kappa-3/2}\right) = \beta_{c,\parallel} \,\left(\frac{\kappa}{\kappa-3/2}\right)\,,
\label{eq:beta}
\end{equation}
with $\beta_{\parallel} \to \beta_{c,\parallel}$ when $\kappa \to
\infty$. Thus, to make a proper comparison between our cases, here we
need to consider variations of the independent one, i.e.,
$\beta_{c,\parallel}$, associated with the core.
\begin{figure}[t!] 
  \centering
    \includegraphics[width=0.80\textwidth]{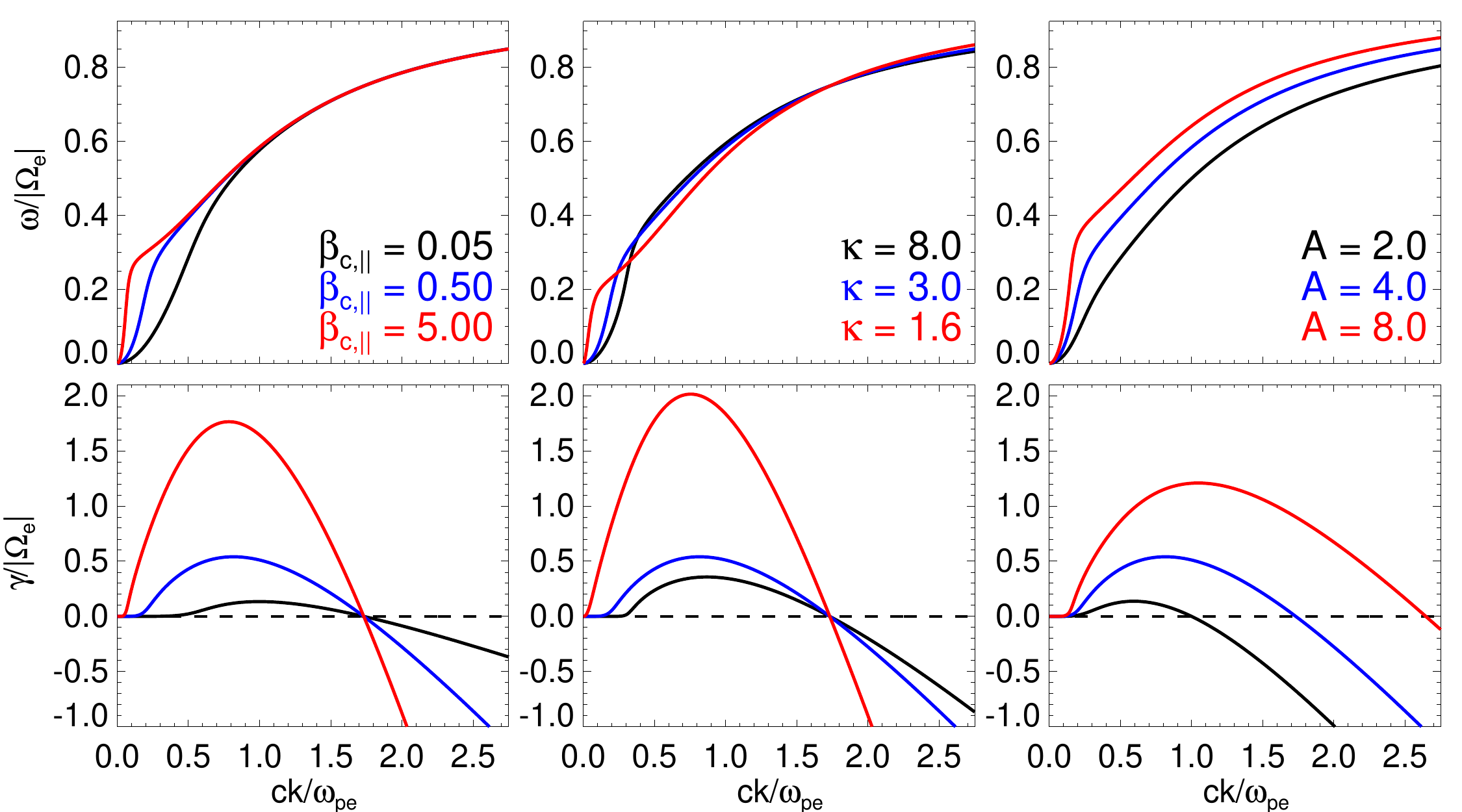}
    \caption{\label{fig:disprel} Dispersion relation for Case I
      (left), Case II (center), and Case III (right) with plasma
      parameters as shown in Table~\ref{t1}. Top and bottom panels
      show the real and imaginary parts of the frequency as a function
      of wavenumber, expressed in units of the electron gyrofrequency
      and inertial length, respectively.}
\end{figure}

Fig.~\ref{fig:disprel} shows the wave-number linear dispersion of the
wave (real) frequency, $\omega$, and the growth rate (imaginary),
$\gamma$, for fixed $\kappa$ and anisotropy, and varying
$\beta_{c,\parallel}$ (Case I, left); fixed anisotropy and
$\beta_{c,\parallel}$, and varying $\kappa$ (Case II, center); and
fixed $\kappa$ and $\beta_{c,\parallel}$ with different anisotropy
values (Case III, right), considering plasma parameters detailed in
Table \ref{t1}. With this systematic variation of the three main
parameters controlling the dispersion relation is clear that when all
other parameters are fixed, $\gamma$ increases with increasing
$\beta_{c,\parallel}$, $A$, and decreasing $\kappa$ (see bottom panels
in the figure). Thus, as during QL relaxation the energy of the
electromagnetic waves grows at the expense of temperature anisotropy
(which means also variations of the plasma beta), if the value of
$\kappa$ also determines the level of the instability, it is expected
$\kappa$ to also evolve as the instability develops. In other words,
if the anisotropy and beta parameters are not constant during the
evolution of the plasma, then $\kappa$ should not remain constant
either. However, as QL evolution of the EMEC instability produces a
negative variation of the anisotropy ($dA/dt <0$) and a positive
variation of parallel beta ($d\beta_\parallel/dt <0$), the sign of
$d\kappa /dt$ remains still unclear. We will address this topic in the
next section.

\section{Quasi-linear evolution with a dynamic {\large $\kappa$}}
\label{sec:qlt}

To find a consistent model to account for a QL relaxation with
$d\kappa /dt \neq 0$ we start from the traditional QL
approach. Following Lazar et al. (2017)~\cite{Lazar2017}, using QL
theory we find the equations describing the EMEC instability that are
given by the time derivatives of both temperatures
\begin{eqnarray}
\frac{dT_\perp}{dt} &=& -\frac{e^2}{m_e k_B}\int_{-\infty}^\infty \frac{dk_\parallel}{c^2k_\parallel^2}\,\omega_i(k_\parallel)
\left(\frac{c^2k_\parallel^2}{\omega_{pe}^2} +\frac{1}{2}\right)\delta B^2(k_\parallel), \label{eq:dtperp} \\
\frac{dT_\parallel}{dt} &=& \frac{e^2}{m_e k_B} \int_{-\infty}^\infty\frac{dk_\parallel}{c^2k_\parallel^2}
\,\omega_i(k_\parallel)\left(\frac{c^2k_\parallel^2} {\omega_{pe}^2}+1\right)\delta B^2(k_\parallel), \label{eq:dtpara}
\end{eqnarray}
and the wave energy density equation
\begin{equation}
\frac{\partial\delta B^2(k_\parallel)}{\partial t} =2\omega_i(k_\parallel)\,\delta B^2(k_\parallel). \label{eq:bk2}
\end{equation}

In the case of bi-Maxwellian distributions,
Eqs.~(\ref{eq:dtperp})-(\ref{eq:bk2}) together with the dispersion
relation Eq.~(\ref{eq:disprel}) provide a closed system of
differential equations. However, in the case of bi-Kappa distribution,
from the expressions of temperatures in Eqs.~(\ref{eq:tthkappa}),
their variation in time, i.e., $dT_{\parallel, \perp}/dt$ implies
variations in time not only for $\Theta_{\parallel, \perp}$ (as
assumed in a simplified model with fixed $\kappa$) but also for the
exponent $\kappa$
\begin{equation}
 \frac{3}{2 \kappa (\kappa - 3/2)} \frac{d \kappa}{d t}=  \frac{1}{\Theta} \frac{d \Theta}{d t} - \frac{1}{T} \frac{d T}{d t}\,. \label{eq:dkdt}
\end{equation}
$\kappa$ couples both time variations of the temperature components,
such that, at least at $t=0$, the anisotropy of bi-Kappa distribution
is the same with the anisotropy of bi-Maxwellian core, as shown in
Eq.~(\ref{eq:aniso}). These variations can be studied for two distinct
cases. First, if $\kappa = \kappa_0 = $ constant, the relaxation of
the temperature anisotropy is achieved by $d T_{\perp} \leq 0$ and $d
T_{\parallel} \geq 0$, such that $dA/dt <0$, which implies that $d
\Theta_{\perp} \leq 0$ and $d \Theta_{\parallel} \geq 0$~\cite[see
  e.g][]{Lazar2017,Lazar2018,Lazar2019}.
On the other hand, if $\kappa$ changes in time, i.e., $\kappa =
\kappa(t)$ (an evolution suggested by the Vlasov simulations in
Ref.~\cite{Lazar2017}), the QL evolution of the growing fluctuations
involves the relaxation of bi-Kappa VDF and is described by the time
variation of seven variables, i.e., $T_{\perp,\parallel}$, $\delta B$,
$\omega_i$, $\kappa$, and also $\Theta_{\perp,\parallel}$. However, we
only have five equations of evolution, i.e.,
Eqs.~(\ref{eq:dtperp})-(\ref{eq:dkdt}), and the dispersion relation
Eq.~(\ref{eq:disprel}), and need two more independent equations, e.g.,
for the evolution of the core temperatures
$\Theta_{\perp,\parallel}$. Here we propose a zero-order approximation
which assumes the dynamics dominated by the quasi-thermal core, with
an evolution described (independently of suprathermals) by the
following QL equations
\begin{eqnarray}
\frac{d\Theta_\perp}{dt} &=& -\frac{e^2}{m_e k_B}\int_{-\infty}^\infty \frac{dk_\parallel}{c^2k_\parallel^2}\,\omega_i^c(k_\parallel)
\left(\frac{c^2k_\parallel^2}{\omega_{pe}^2} +\frac{1}{2}\right)\delta B_c^2(k_\parallel), \label{eq:dThetaperp} \\
\frac{d\Theta_\parallel}{dt} &=& \frac{e^2}{m_e k_B} \int_{-\infty}^\infty\frac{dk_\parallel}{c^2k_\parallel^2}
\,\omega_i^c(k_\parallel)\left(\frac{c^2k_\parallel^2} {\omega_{pe}^2}+1\right)\delta B_c^2(k_\parallel). \label{eq:dThetapara}
\end{eqnarray}
Note that Eqs.~(\ref{eq:dThetaperp})-(\ref{eq:dThetapara}) are the
correspondent equations to Eqs.~(\ref{eq:dtperp})-(\ref{eq:dtpara}),
but instead of the evolution of the bi-Kappa describe the evolution of
bi-Maxwellian core ($f_M$) under the effects of whistler growing
electromagnetic fields (EMEC instability)
\begin{equation}
     \frac{\partial\delta B_c^2(k_\parallel)}{\partial t} =2\omega_i^c(k_\parallel)\,\delta B_c^2(k_\parallel), \label{eq:bk2core}
\end{equation}
triggered only by the anisotropic core with growth rate $\omega_i^c$ given by the dispersion relation
\begin{equation}
     \frac{c^2k_\parallel^2}{\omega_{pe}^2} = \left(\frac{\Theta_\perp}{\Theta_\parallel}-1\right)+
\left[\frac{\Theta_\perp}{\Theta_\parallel}\,\omega -\left(\frac{\Theta_\perp}{\Theta_\parallel}-
1\right)|\Omega_e|\right] \frac{1}{k_\parallel\sqrt{2\Theta_\parallel /m}}\,Z
\left(\frac{\omega-|\Omega_e|}{k_\parallel\sqrt{2\Theta_\parallel /m}}\right), \label{eq:disprelCore}
\end{equation}
where $Z(x)$ is the standard plasma dispersion function characteristic
to bi-Maxwellian plasmas, given by Eq.~(\ref{eq:z}). As already
mentioned, due to its high density ($>$ 90\% of total density) we
assume that the core dominates the dynamics, such that in this zero
order approximation we can decouple the evolution of the core from the
evolution of suprathermal tails, and implicitly from the variation of
kappa index. In summary,
Eqs.~(\ref{eq:dThetaperp})-(\ref{eq:dThetapara}), completed by
Eqs. (\ref{eq:bk2core})-(\ref{eq:disprelCore}), describe the time
variation of the core, i.e., moments $\Theta_{\perp, \parallel}$,
which can be then coupled to Eqs.~(\ref{eq:dtperp})-(\ref{eq:dkdt})
and (\ref{eq:disprel}) to model time variation of $\kappa$ parameter
under QL theory.

\section{Results}
\label{sec:results}

To solve the system of QL equations, we use a discrete grid in the
wavenumber space (normalized to the electron inertial length) with
2000 points between $0.001 < |c k_\parallel/\omega_{pe} < 3$. All
cases were ran up to $\Omega_e t = 300$ with a time step of $dt =
0.02/\Omega_e$, and the magnetic field spectrum was chose to be
$\delta B^2(k_\parallel) = 10^{-5} B^2_0$ at $t=0$. Under this
procedure, knowing the magnetic field spectrum and the value of the
parameters at any given time $t$, we can solve the dispersion relation
to find the complex frequency $\omega+i\gamma$, as a function of
$k_\parallel$ at this particular time, to then evaluate the time
derivative of each parameter. We then evolve the whole system to the
next time step $t+dt$ with a 2nd order Runge-Kutta method. Using this
scheme we solved the QL system for all 9 cases shown in Table~\ref{t1}
as initial conditions.

\subsection{Case I. Varying plasma beta}
\label{sec:caseI}

\begin{figure}[t!] 
  \centering
    \includegraphics[width=0.80\textwidth]{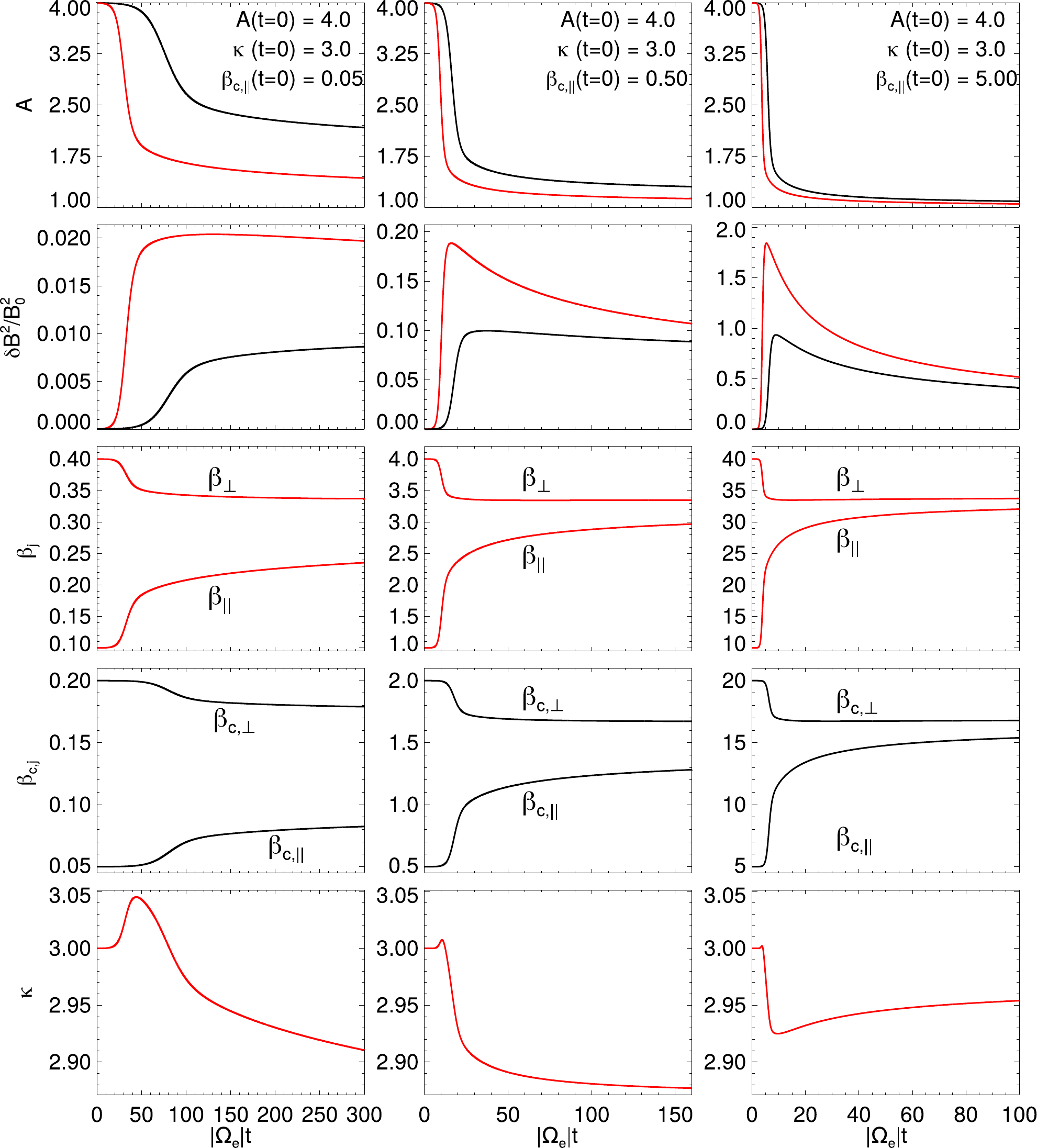}
    \caption{\label{fig:caseI} QL time evolution of the EMEC
      instability considering initial parameters from Table~\ref{t1}
      Case I, fixing $\kappa(t=0) = 3.0$, and $A(t=0) = 4.0$ and
      varying $\beta_{c,\parallel}=0.05$ (left), $\beta_{c,
        \parallel}=0.5$ (center), and $\beta_{c,\parallel}=5.0$
      (right). In all panels, red and black curves represent the
      evolution of the magnetic field and plasma parameters of the
      bi-Kappa or bi-Maxwellian core VDFs, respectively. From top to
      bottom each row shows temperature anisotropy, magnetic energy
      density, parallel and perpendicular plasma of the bi-Kappa,
      betas of the bi-Maxwellian core, and $\kappa$ parameter. In all
      panels, time is expressed in units of the electron
      gyrofrequency.  }
\end{figure}

First we consider the QL evolution of three initial conditions shown
in Table~\ref{t1} Case I, fixing temperature anisotropy and $\kappa$
parameter, and varying plasma beta. In Figure~\ref{fig:caseI} left,
center and right columns correspond to $\beta_{c,
  \parallel}(t=0)=0.05$, $0.5$ and $5.0$, respectively, $\kappa(t=0) =
3$, and $A(t=0) = 4$. In addition, with this selection of parameters
we also have $\beta_{\parallel}(t=0)=0.1$ (left),
$\beta_{\parallel}(t=0)=1.0$ (center), and
$\beta_{\parallel}(t=0)=10.0$ (right). From top to bottom each row
shows the time evolution of temperature anisotropy, the magnetic field
energy, parallel and perpendicular temperature of the bi-Kappa,
parallel and perpendicular temperature of the bi-Maxwellian core, and
$\kappa$ parameter. Black and red curves represent the quantities
associated with the bi-Maxwellian and bi-Kappa VDFs and the associated
magnetic field energy. From the figure we can see that, as expected
for the bi-Maxwellian core and the bi-Kappa, the decrease of
temperature anisotropy (first row) drives the growth of the
fluctuating fields (second row) while temperatures approach each other
(third and fourth rows).

The main differences between each case (column) lie in the timescale
of the process and the behavior of the $\kappa$ parameter (bottom row
in Figure~\ref{fig:caseI}). From left to right each column shows a run
with a larger initial growth rate, and therefore a shorter
characteristic time for the QL relaxation (note that owing to this the
extent of the time axis is different at each column). Even though in
all three cases $\kappa$ increases at the beginning of the relaxation
process, the time profile of $\kappa$ appears to be dependent on the
initial value of beta. For $\beta_\parallel=10.0$ (right column),
initially $\kappa$ exhibits a minimal increase. Then, following the
fast relaxation of the temperature anisotropy, $\kappa$ rapidly
decreases reaching a minimum value at the same time in which
perpendicular temperatures finish their rapid decrease ($\Omega_e
t\sim 9$), and finally slowly recovers to a level slightly higher. For
$\beta_\parallel=1.0$, however, the center column of
Figure~\ref{fig:caseI} shows a different behavior. In this case the
initial increase of $\kappa$, while still small, is about ten times
larger compared to the previous case, and the later decrease of the
$\kappa$ parameter remains monotonous until the end of the
calculations. For $\beta_\parallel=0.1$ (left column in the figure)
the initial increase of $\kappa$ is about ten time larger compared
with the $\beta_\parallel=1.0$ case. In addition, here the changes
occur at a considerably slower rate, but the final value of $\kappa$
is similar.

\subsection{Case II. Varying the $\kappa$ parameter.}
\label{sec:caseII}

After analyzing the 3 cases shown in Figure~\ref{fig:caseI} we can see
that the time evolution of the $\kappa$ parameter is not
trivial. $\kappa(t)$ is not even a monotonously increasing or
decreasing function of time as the anisotropy or the
temperatures. Here we continue the analysis by considering Case II
initial values from Table~\ref{t1}, fixing parallel beta of the
Maxwellian core and temperature anisotropy at $t=0$ to
$\beta_{c,\parallel}=0.5$, and $A= 4$, respectively, and selecting
three different initial values for $\kappa$. Namely, $\kappa =1.6$
(left), $\kappa=3.0$ (center) and $\kappa=8.0$ (right), which implies
$\beta_\parallel=8.0$, $1.0$, and $0.62$, respectively.
\begin{figure}[t!] 
  \centering
    \includegraphics[width=0.80\textwidth]{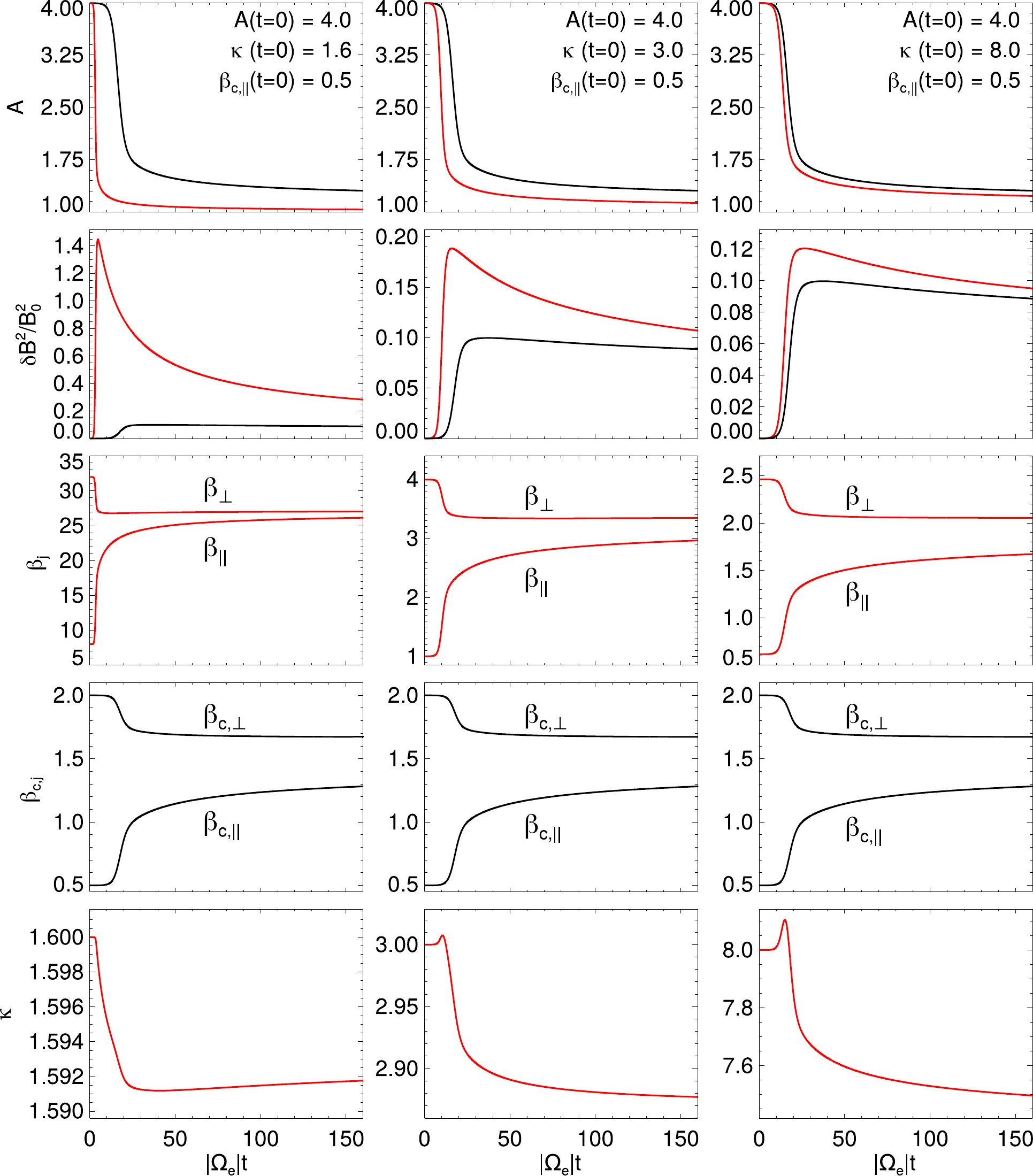}
    \caption{\label{fig:caseII} QL time evolution of the EMEC
      instability considering initial parameters from Table~\ref{t1}
      Case II, fixing $A(t=0) = 4.0$, and $\beta_{c,\parallel}(t=0) =
      0.5$ and varying $\kappa(t=0)=1.6$ (left), $\kappa(t=0)=3.0$
      (center), and $\kappa(t=0)=8.0$ (right), which implies
      $\beta_\parallel=8.0$, $1.0$, and $0.62$, respectively. In all
      panels, red and black curves represent the evolution of the
      magnetic field and plasma parameters of the bi-Kappa or
      bi-Maxwellian core VDFs, respectively. From top to bottom each
      row shows temperature anisotropy, magnetic energy density,
      parallel and perpendicular plasma of the bi-Kappa, betas of the
      bi-Maxwellian core, and $\kappa$ parameter. In all panels, time
      is expressed in units of the electron gyrofrequency.  }
\end{figure}
 
 The QL evolution for this three cases are shown in
 Figure~\ref{fig:caseII} where each panel represent the same as in
 Figure~\ref{fig:caseI}. These three cases with different initial
 $\kappa$ parameter (three columns in Figure~\ref{fig:caseII}) exhibit
 qualitatively the same time evolution, but all occurring faster and
 with a larger level of electromagnetic fluctuations for smaller
 initial $\kappa$ (larger initial $\beta_\parallel$. This is
 consistent with the situation shown in Figure~\ref{fig:disprel}, as a
 smaller value of $\kappa$ implies a larger growth rate and therefore
 a faster relaxation. In addition, from Figure~\ref{fig:caseII} we
 also see that the absolute variation of $\kappa$ seem to scale with
 the value of $\kappa$, in line with Eq.~(\ref{eq:dkdt}). As
 $d\kappa/dt$ is proportional to $\kappa -3/2$, independent of the
 value of other plasma parameters, a larger initial value of $\kappa$
 will always yield to a larger time variation of the same
 parameter. Finally, Figure~\ref{fig:caseII} also shows that, even
 though the final value of $\kappa$ is smaller than $\kappa(t=0)$, the
 $\kappa$ parameter increases at the beginning of the relaxation, and
 the initial enhancement of $\kappa$ increases with increasing initial
 $\kappa$ value.

\subsection{Case III. Varying temperature anisotropy.}
\label{sec:caseIII}

\begin{figure}[t!] 
  \centering
    \includegraphics[width=0.8\textwidth]{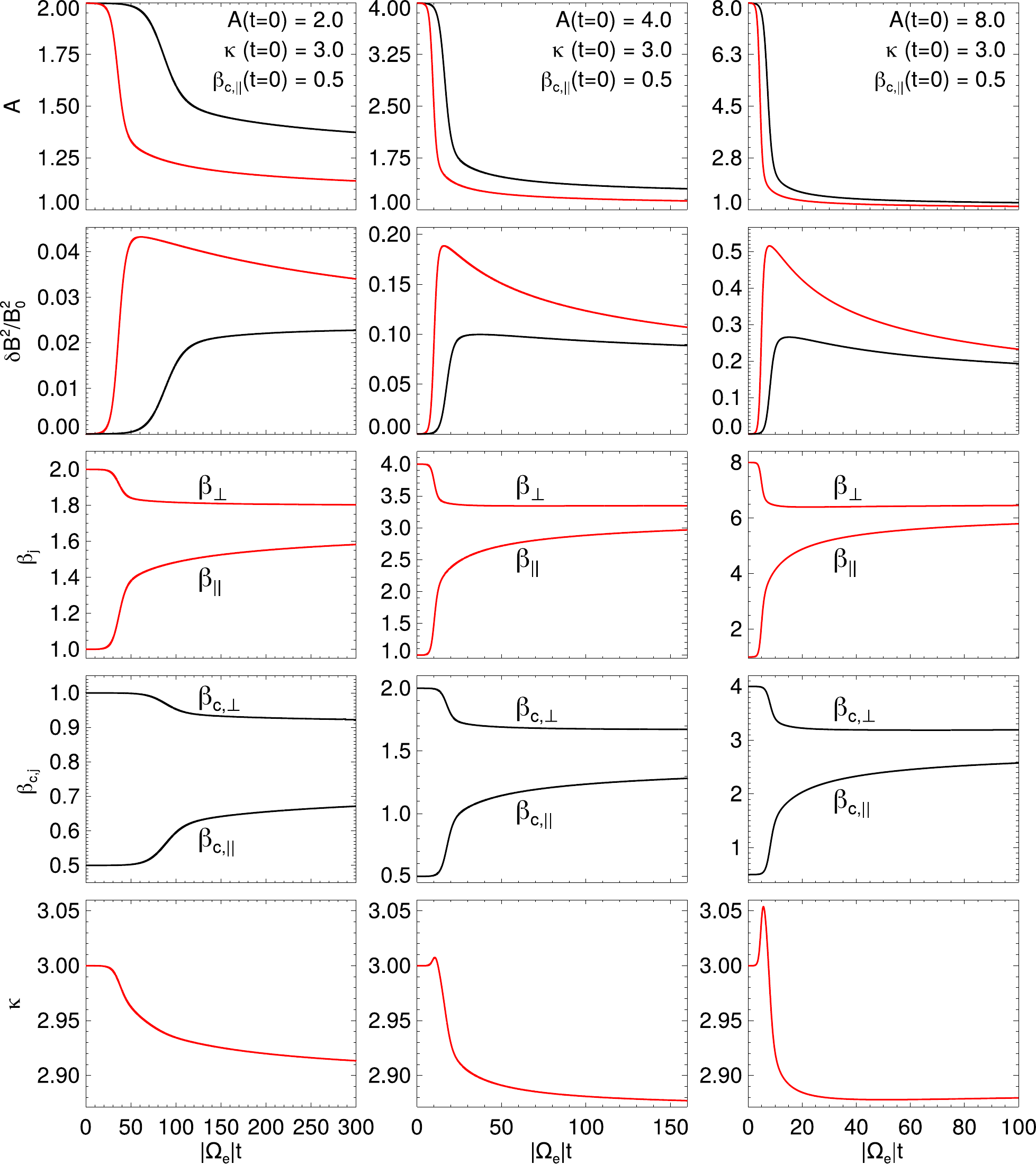}
    \caption{\label{fig:caseIII} QL time evolution of the EMEC
      instability considering initial parameters from Table~\ref{t1}
      Case III, fixing $\kappa(t=0) = 3.0$, and
      $\beta_{c,\parallel}(t=0) = 0.5$ ($\beta_\parallel(t=0) = 1.0$),
      and varying $A(t=0)=2.0$ (left), $A(t=0)=4.0$ (center), and
      $A(t=0)=8.0$ (right). In all panels, red and black curves
      represent the evolution of the magnetic field and plasma
      parameters of the bi-Kappa or bi-Maxwellian core VDFs,
      respectively. From top to bottom each row shows temperature
      anisotropy, magnetic energy density, parallel and perpendicular
      plasma of the bi-Kappa, betas of the bi-Maxwellian core, and
      $\kappa$ parameter. In all panels, time is expressed in units of
      the electron gyrofrequency.  }
\end{figure}

To further characterize the evolution of $\kappa$ depending on the
initial parameters we now fix $\beta_{c,\parallel}=0.5$ and $\kappa
=3.0$ at $t=0$ (implying $\beta_\parallel(t=0)=1.0$, and consider
three possibilities for the initial temperature
anisotropy. Figure~\ref{fig:caseIII} shows the QL evolution for
$A=2.0$ (left), $A=4.0$ (center), and $A = 8.0$ (right) at $t=0$, with
all other initial plasma parameters given by Case III in
Table~\ref{t1}. From the figure we can see that this case is similar
to the previous ones, but as expected from linear calculations here
the initial anisotropy determines the time scale of the dynamics. The
figure also shows that the initial temperature anisotropy also
controls the shape of $\kappa$ as a function of time. In the same
fashion as the cases shown in Figure~\ref{fig:caseI} and
Figure~\ref{fig:caseII}, for $A(t=0) =8.0$ and $A(t=0) =4.0$ the
$\kappa$ parameter initially increases and then decreases, and the
level of the initial increase depend on how fast the relaxation
occurs. However, for $A(t=0) = 2.0$ the $\kappa$ parameter only
decreases with time. Thus, unlike Figure~\ref{fig:caseI} in which a
faster relaxation resulted in a smaller increase of $\kappa$, when the
speed of the relaxation is determined by the initial temperature
anisotropy, $\kappa$ reaches larger values for larger growth rate of
the waves.

After considering all these cases a robust emerging result is that,
regardless the initial value of the anisotropy, $\Delta \kappa =
\kappa(t_{\rm{end}}) - \kappa(t=0) \leq 0$, where
$\kappa(t_{\rm{end}})$ is the value of $\kappa$ after the relaxation
of the EMEC instability; i.e., independent of the details of each
case, the QL relaxation of the EMEC instability produces an absolute
decrease of $\kappa$. It is important to mention that this is
consistent with Vlasov-Maxwell simulations presented in
Ref.~\cite{Lazar2017}, in which, for almost all considered cases the
triggering and relaxation of the EMEC instability led to a reduction
of the $\kappa$ parameter in a qualitatively similar timescale.

\subsection{Plasma beta and temperature anisotropy paths}

We have already solved the equations of our QL model considering
distinct combinations of the relevant initial plasma parameters
(temperature anisotropy, plasma beta and $\kappa$). With such
systematic study we were able to identify and characterize the effect
of each parameter on the QL evolution of the EMEC instability. Here we
complete our analysis by performing an ensemble of QL runs for
different values of the initial $\kappa$ parameter ($1.6 \leq
\kappa(t=0) \leq 10.0$ and a parallel beta ($<0.1 \leq
\beta_\parallel(t=0) \leq 10.0$), and also fixing $A(t=0) = 4.0$. With
this procedure we can build paths in the $\beta_\parallel$, $A$
parameters space, and follow the dynamics of the EMEC instability
triggered from different starting points at $t=0$ up to $|\Omega_e| t
= 300$, which is a long enough time scale for the development of the
EMEC instability for all considered initial conditions.
\begin{figure*}[t!] 
  \centering
\includegraphics[width=0.95\textwidth]{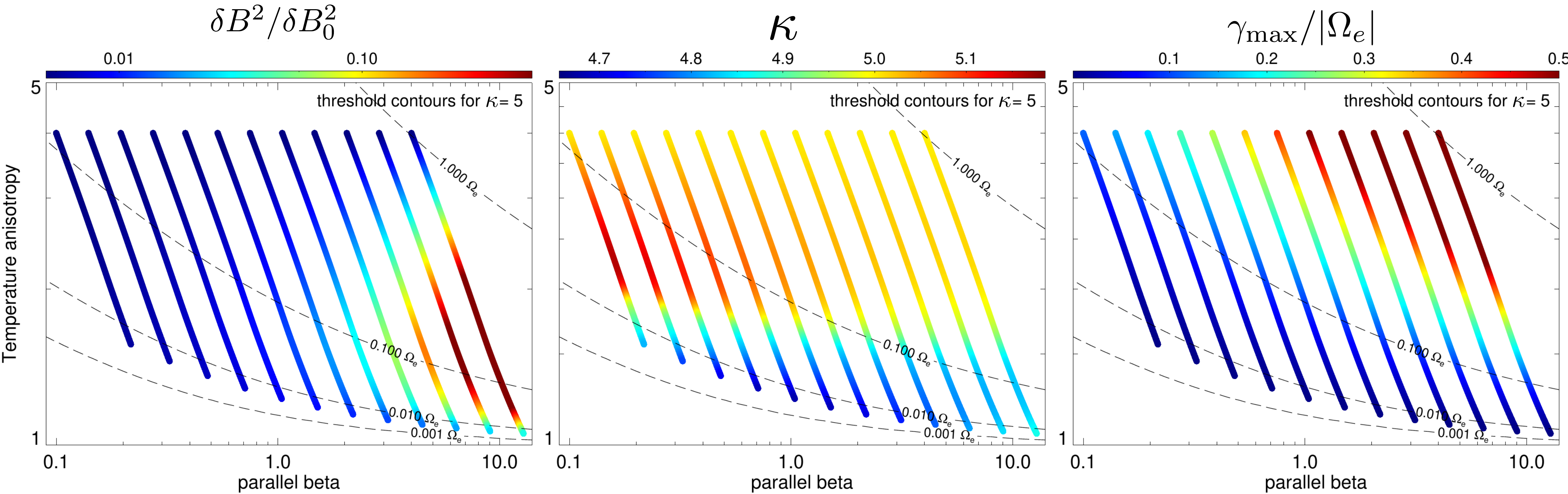}
    \caption{\label{fig:paths}Time evolution in the
      ($\beta_\parallel$, $A$)-space, from an ensemble of QL
      calculations, starting from the same initial $A(t=0) = 4$ and
      $\kappa_0 = 5$, up to $|\Omega_e|t = 300$), for the fluctuating
      magnetic energy (left), $\kappa$ (middle) and $\gamma_{\rm max}$
      (right). Dashed lines mark different maximum growth rate
      contours.  }
\end{figure*}

Figure~\ref{fig:paths} shows dynamical paths of the magnetic field
energy (left panel), the $\kappa$ parameter (center), and the maximum
growth rate of EMEC waves (right), during the relaxation of the EMEC
instability for $\kappa(t=0) = 5.0$ (other values of initial $\kappa$
exhibit similar results and therefore are not shown here). Regarding
the magnetic field energy (left panel in the figure), as expected, the
level of the electromagnetic fluctuations increases with increasing
initial parallel beta. At the same time, the plasma evolves towards a
less unstable state by reducing temperature anisotropy and increasing
parallel plasma beta, with all dynamical paths collapsing to a EMEC
instability level with $10^{-3} \Omega_e < \gamma < 10^{-2} \Omega_e$
(dashed lines in the figure). In addition, looking a the evolution of
$\kappa$ (central panel in Figure~\ref{fig:paths}), we can see that in
all cases the final value of $\kappa$ is less than the initial one
($\Delta \kappa < 0$). At the same time, in all cases the maximum
growth rate monotonically decreases in time (see right panel in
Figure~\ref{fig:paths}). Even though, as for fixed plasma beta and
anisotropy the growth rate of the waves increases with decreasing
$\kappa$, when the $\kappa$ parameter is allowed to change in time,
the relaxation or saturation of the EMEC instability is achieved not
only by the relaxation of the temperature anisotropy (as in the case
of a bi-Maxwellian) but also by the decrease of the $\kappa$
parameter. Therefore, on the contrary to the case of bi-Maxwellian
distributions, these results seem to indicate that for bi-Kappa
distributions, the collisionless relaxation of the EMEC instability
does not result in the thermalization of the plasma as such behavior
would imply $\Delta \kappa > 0$. Instead of thermodynamic equilibrium,
the plasma does evolve towards stability, but in this case stability
according to the Vlasov equation.

\begin{figure}[ht!] 
  \centering
   \includegraphics[width=0.7\textwidth]{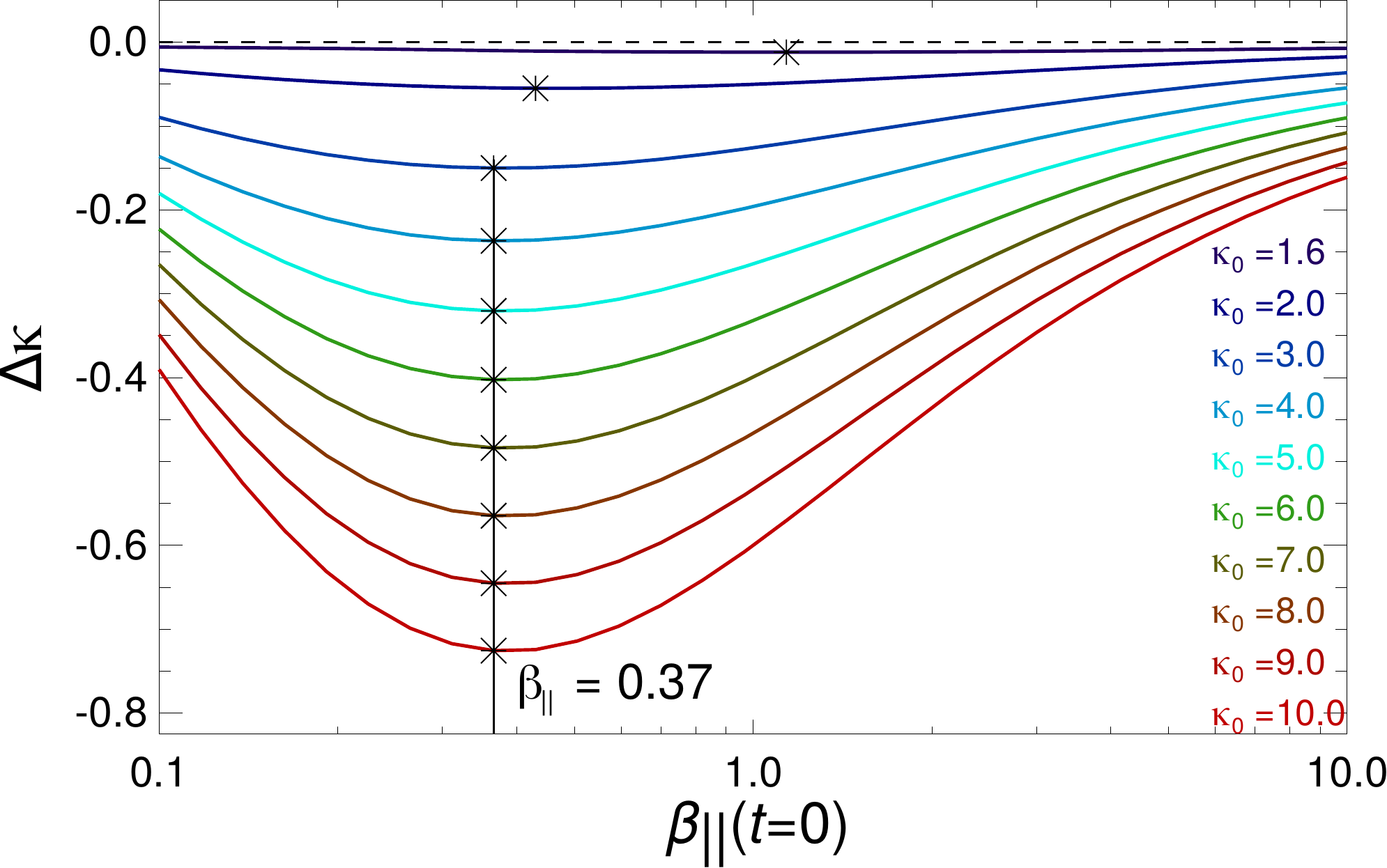}
    \caption{\label{fig:deltakappa} Difference between the value
      $\kappa$ after and before the QL relaxation as a function of
      initial parallel plasma beta. Each color correspond to a
      different initial $\kappa$ value, and black asterisk symbols
      mark the maximum $\kappa$ variation for each initial
      $\kappa$. For all $\kappa \geq 3.0$ the initial beta allowing
      the largest variation of the $\kappa$ parameter is
      $\beta_{\rm{min}}\sim 0.37$.  }
\end{figure}

Finally, although for all cases $\Delta \kappa < 0$, as already
mentioned, the evolution and relaxation of the $\kappa$ parameter is
considerably more complex than the relaxation of the temperature
anisotropy. A closer look at the central panel of
Figure~\ref{fig:paths} suggests that for $\kappa(t=0) = 5.0$ there is
a range of initial plasma beta values around $\beta_\parallel
(t=0)\sim 0.5$, in which the reduction of $\kappa$ is maximum. If beta
is too small $\kappa$ also decreases, but only after an initial
increasing which is bigger for decreasing bet such that the addition
of this two-steps process determines $\Delta \kappa$ to be small. On
the other hand, if $\beta_\parallel (t=0) > 1.0$, then the variation
of $\kappa$ is also smaller. The same situation repeats for all
considered initial $\kappa$ values as shown in
Figure~\ref{fig:deltakappa}. From the figure we can see that for all
considered initial $\kappa$ values, $\Delta \kappa$ is a negative
function of beta with a clear minimum. For small initial $\kappa$
($\kappa(t=0) = 1.6$ or 2.0), the reduction of $\kappa$ is almost
negligible and the maximum reduction of $\kappa$ is achieved with
$\beta_{\rm{min}}\sim 0.5$. However, for all $\kappa \ge 3.0$ the
initial beta allowing the largest variation of the $\kappa$ parameter
is $\beta_{\rm{min}}\sim 0.37$, and that the maximum absolute
variation of $\kappa$ increases with increasing $\kappa$. All these
results confirm that the QL relaxation of the EMEC instability lead
the plasma towards a Vlasov equilibrium (meaning the quasi-stationary
state reached in the absence of collisions) and not necessarily
thermodynamic equilibrium.
\section{Conclusions}
\label{sec:conclusions}

In this work we have introduced a new QL kinetic approach to account
for the variations of all parameters characterizing a bi-Kappa
distribution function during the excitation of whistler
instability. Going one step further than previous moment-based QL
works we have been able to address the variability and evolution of
the $\kappa$ parameter due to wave-particle interactions, and find
that indeed $\kappa$ is a dynamic parameter that can evolve during the
excitation of this instability. A potential variation of $\kappa$ was
already suggested by the fully non-linear
simulations~\cite{Lu2010,Eliasson2015,Lazar2017,Lazar2018},
considering the time evolution of the whistler fluctuations and their
feedback on the bi-Kappa distributed electrons.

Considering plasma parameters relevant to space environments such as
the solar wind or planetary magnetospheres, we have performed a
systematic study to analyze the dependence of the initial value of
temperature anisotropy, plasma beta, and the $\kappa$ parameter. Our
numerical results show that the time variation of $\kappa$ increases
for higher initial values $\kappa(t=0)$, and that the time evolution
of $\kappa(t)$ is not monotonic as in the case of temperature
anisotropy that is constantly decreasing in time (or an increasing
parallel plasma beta). In general, we found $\kappa$ increasing at the
beginning of the QL evolution, and then decreasing towards a value
smaller than the initial one, such that, overall $\Delta \kappa =
\kappa(t_{\rm{end}}) - \kappa(t=0) \leq 0$. As already mentioned, the
plasma beta and anisotropy show different evolutions. The initial
growth of $\kappa$ increases for larger initial temperature
anisotropy, but decreases for decreasing initial parallel beta. In
addition, the final variation of $\kappa$ ($|\Delta \kappa|$) exhibits
a maximum value for initial parallel beta around
$\beta_\parallel\sim0.37$ for all $\kappa(t=0)\ge 3.0$, suggesting
that for extreme values of beta the $\kappa$ power-index may not be
affected by wave-particle interactions.

All these results show a variable $\kappa$ during the excitation of
kinetic instabilities. The QL evolution of the EMEC instability
results in a reduction of $\kappa$, such that plasma evolves towards
Vlasov equilibrium, but not necessarily a thermodynamic (Maxwellian)
equilibrium, which is somehow expected in the presence of a
fluctuating EM power, low but still present in the system (preventing
also the complete relaxation of temperature anisotropy). As shown in
Figs.~\ref{fig:caseI}-\ref{fig:paths}, the QL (partial) relaxation of
the anisotropic temperature leads to an increased level of wave
fluctuations in the plasma, such that the system achieves a new
quasi-stationary (quasi-stable) state with an enhanced level of wave
fluctuations which entertain and boost, evenly, the supathermals,
ultimately lowering the parameter $\kappa$. Throughout all these
processes the electromagnetic turbulence plays an important role on
the suprathermalization of the plasma and may even determine the lower
values of the kappa power-index~\cite[see e.g.][]{Yoon2019}.

Even though with this zero-order model the variations of $\kappa$ can
be considered rather small, the consistency of the results here
presented, and their good qualitative agreement with previous reports
based on fully non-linear models~\cite{Lu2010,Lazar2017} suggests that
QL approaches can be used as a powerful tool to study the non-linear
evolution of poorly collisional plasmas. This is true for
bi-Maxwellian plasmas in which density, temperatures and bulk velocity
describe the dynamics of the system~\cite[see e.g.][and references
  therein]{Yoon2017}, and the same should occur in the case of
non-Maxwellian plasmas exhibiting non-thermal features such as
power-law tails. We plan to address these aspects and other issues,
expanding the scope of our model in subsequent works.

In summary, the fact that Kappa distributions are ubiquitous in many
space and astrophysical plasmas, suggests that the enhanced
fluctuations triggered by kinetic instabilities may play a key role on
the regulation of these distributions in the inner heliosphere.  Under
this context, our new QL approach may improve the understanding of the
processes that control and shape the velocity distribution function of
plasma particles in poorly collisional plasmas. We expect these
results to motivate the community to consider more realistic
theoretical models, which complement the more idealized
(bi-)Maxwellian description with $\kappa$ power exponents, or any
other plasma parameter that can reproduce the observations. Such
models are expected to be validated by the high resolution
observations by the Solar Parker Probe or Solar Orbiter missions,
which can couple particle and wave fluctuations from concomitant
in-situ measurements.

\ack These results were obtained in the framework of the projects SCHL
201/35-1 (DFG-German Research Foundation), C14/19/089 (C1 project
Internal Funds KU Leuven), G.0D07.19N (FWO-Vlaanderen), C~90347 (ESA
Prodex 9), and Fondecyt No. 1191351 (ANID, Chile). We would like to
thank Rodrigo Lopez and Roberto Navarro for useful
discussion. P.S. Moya is also grateful for the support of KU Leuven
BOF Network Fellowship NF/19/001.

\bibliographystyle{iopart-num}
\bibliography{biblio}
 
\end{document}